\documentclass{ws-ijmpcs}

\usepackage{amsmath,bm}
\usepackage{graphicx}
\usepackage{hyperref}

\begin{document}

\markboth{K. V. Shajesh and M. Schaden}
{Significance of many-body contributions to Casimir energies}

\catchline{}{}{}{}{}
%
%%%%%%%%%%%%%%%%%%%%%%%%%%%%%%%%%%%%%%%%%%%%%%%%%%%%%%%%%%%%%%%%%%%%

\title{Significance of many-body contributions to Casimir energies}
\author{K. V. Shajesh and M. Schaden}
%\homepage{http://andromeda.rutgers.edu/~shajesh}
\address{Department of Physics, Rutgers, The State University of New Jersey, \\
101 Warren Street, Newark, NJ - 07102, USA. \\
shajesh@andromeda.rutgers.edu
}

\maketitle

\begin{history}
%Last compiled on {\today} \\
\received{Day Month Year}
\revised{Day Month Year}
\end{history}

%--------------------------------------------
\begin{abstract}
Irreducible many-body contributions to Green's functions and 
Casimir energies are defined. We show that the irreducible
three-body contribution to Casimir energies are significant
and can be more than twenty percent of the total interaction energy.
Irreducible three-body contribution for three parallel semitransparent 
plates in the limit when two plates overlap is obtained in terms of 
irreducible two-body contributions and shown to be finite and well defined
in this limit.
\end{abstract}

\keywords{Many-body Casimir energies, Many-body Green's functions, 
Faddeev's equations.}

\ccode{PACS numbers: 11.10.-z, 11.10.Jj, 11.80.-m, 11.80.Jy, 11.80.La}

%--------------------------------------------
%\tableofcontents
%--------------------------------------------

\section{Introduction}

Three-body contribution to Casimir energies in the unretarded regime
were first considered by Axilrod and Teller\cite{Axilrod:1943at},
and Muto\cite{Muto:1943fc}. The three-body contribution to 
van der Waals-London interaction energy between three identical atoms
at the corners of a triangle was found\cite{Axilrod:1943at}
to be negative for configurations
forming an acute triangle and positive for configurations forming 
obtuse triangles. Axilrod\cite{Axilrod:1949gr,Axilrod:1951ti,Axilrod:1951gr}
hoped to explain the cohesion energy of rare-gas crystals
by including three-body contributions. In the context of atoms
many-body contributions to Casimir-Polder interaction energy in the
retarded regime were studied by Aub and Zienau\cite{Aub:1960a}.
Three-body contribution to the Casimir-Polder 
interaction of two spheres above a plate was recently explored in 
Refs.~\refcite{Rodriguez:2009fn} and \refcite{Varela:2011fe}.
Due to their non-additive nature, irreducible many-body contributions to 
the total Casimir energy in the strong coupling regime were only
considered recently\cite{Schaden:2010wv,Shajesh:2011ec}.
Theorems on finiteness of irreducible many-body contributions to
Casimir energies were obtained in Ref.~\refcite{Schaden:2010wv},
and for scalar fields with potential interactions the sign of 
irreducible many-body contributions was determined\cite{Schaden:2010wv}.
Explicit expressions for many-body contributions 
to Casimir energies were derived in Ref.~\refcite{Shajesh:2011ec}.
This used ideas of Faddeev and others%
\cite{francis:1953a,brueckner:1953a,brueckner:1954a,Faddeev:1965a}
to solve the many-body Green's functions\cite{Shajesh:2011ec}.
Significance of many-body contributions to 
Casimir energies becomes apparent by noting that the three-body 
contribution can be up to 20\% of the total Casimir energy.
The importance of such non-additive interactions has been
realized by chemists\cite{Chakravarty:1997tg}.

We here first review our results on many-body Green's functions
in Ref.~\refcite{Shajesh:2011ec} and consider some 
implications of the many-body decomposition of Casimir energies.
For (scalar) atom-like potentials above a Dirichlet plate we
analytically obtain the three-body contribution to the Casimir force.
We consider weakly interacting wedges placed
atop Dirichlet plates and show that the irreducible three-body Casimir 
energy is minimal (and vanishes) when the shorter side of the wedge is
perpendicular to the Dirichlet plate.

%--------------------------------------------

\section{Many-body Green's functions}

The (scalar) Green's function for $N$ potentials
$V_i({\bf x})$, $i=1,2,\ldots,N$ satisfies the equation
\begin{equation}
\Big[-{\bm\nabla}^2 +\zeta^2
+ V_1({\bf x}) + V_2({\bf x}) +\ldots + V_N({\bf x}) 
\Big]G_{1\ldots N}({\bf x},{\bf x}^\prime) 
=\delta^{(3)}({\bf x}-{\bf x}^\prime).
\label{Ngeqn-def}
\end{equation}
The solution is symbolically written in the form
\begin{equation}
G_{1\ldots N}=G_0- G_0 T_{1\ldots N}G_0,
\end{equation}
where the free Green's function $G_0({\bf x},{\bf x}^\prime)$
satisfies Eq.~(\ref{Ngeqn-def})
in the absence of potentials and the $N$-body transition matrix 
$T_{1\ldots N}\to T_{1\ldots N}({\bf x},{\bf x}^\prime)$
is of the form
\begin{equation}
T_{1\ldots N} = (V_1+V_2+\dots +V_N)
\Big[ 1 + G_0 (V_1+ V_2+\dots + V_N)\Big]^{-1}.
\end{equation}
To compactly express the following equations we use the notation
\begin{equation}
\tilde G_i\to G_iG_0^{-1}, \qquad
\tilde V_i\to G_0V_i, \qquad \text{and} \qquad \tilde T_i\to G_0T_i,
\end{equation}
which is equivalent to setting $G_0=1$.
We decompose the $N$-body transition matrix in the form,
\begin{equation}
\tilde T_{1\ldots N} = \sum_{i=1}^N\sum_{j=1}^N \tilde T_{1\ldots N}^{ij}
=\text{Sum} \big[ \tilde{\bf T}_{1\ldots N} \big],
\label{T1N-deqn}
\end{equation}
where the symbol $\text{Sum}[{\bf A}]$ stands for the sum of all
elements of the matrix ${\bf A}$.
The matrix form of the $N$-body transition operator is,
\begin{equation}
\tilde{\bf T}_{1\dots N} = \left(
\begin{array}{cccc}
\tilde T_{1\dots N}^{11} & \tilde T_{1\dots N}^{12} 
& \cdots & \tilde T_{1\dots N}^{1N} \\[2mm]
\tilde T_{1\dots N}^{21} & \tilde T_{1\dots N}^{22} 
& \cdots & \tilde T_{1\dots N}^{2N} \\[2mm]
\vdots &\vdots &\ddots & \vdots \\[2mm]
\tilde T_{1\dots N}^{N1} & \tilde T_{1\dots N}^{N2} 
& \cdots & \tilde T_{1\dots N}^{NN}
\end{array} \right),
\end{equation}
where each component is an integral operator.
It was shown\cite{Shajesh:2011ec} that the above $N$-body transition 
matrices satisfy the Faddeev's 
equations\cite{Faddeev:1965a,francis:1953a,brueckner:1954a}
\begin{equation}
\big[{\bf 1} +\tilde{\bm \Theta}_{1\ldots N} \big]
\cdot\tilde{\bf T}_{1\dots N}= \tilde{\bf T}_\text{diag},
\label{fadeqn}
\end{equation}
where
\begin{equation}
\tilde {\bm \Theta}_{1\ldots N} = \left(
\begin{array}{ccccc}
0 & \tilde T_1 & \tilde T_1 & \cdots & \tilde T_1 \\[1mm]
\tilde T_2 & 0 & \tilde T_2 & \cdots & \tilde T_2 \\[1mm]
\tilde T_3 & \tilde T_3 & 0 & \cdots & \tilde T_3 \\[1mm]
\vdots &\vdots &\vdots &\ddots & \vdots \\[1mm]
\tilde T_N & \tilde T_N & \tilde T_N & \cdots & 0
\end{array} \right),
\qquad \qquad
\tilde {\bf T}_\text{diag} = \left(
\begin{array}{ccccc}
\tilde T_1 & 0 & 0 & \cdots & 0 \\[1mm]
0 & \tilde T_2 & 0 & \cdots & 0 \\[1mm]
0 & 0 & \tilde T_3 & \cdots & 0 \\[1mm]
\vdots &\vdots &\vdots &\ddots & \vdots \\[1mm]
0 & 0 & 0 & \cdots & \tilde T_N 
\end{array} \right).
\end{equation}
Faddeev's equations of Eq.~(\ref{fadeqn}) reduce the problem 
of solving Eq.~(\ref{T1N-deqn}) for the $N$-body transition matrix
to that of inverting 
$\big[{\bf 1}+\tilde{\bf \Theta}_{1\ldots N} \big]$
by solving a set of $N$ linear integral equations.
Remarkably, $\tilde {\bf \Theta}_{1\ldots N}$ depends only on
single-body transition operators. 
The norm of $\tilde {\bf \Theta}_{1\ldots N}$ is less than unity 
(because the norm of single-body transition matrices is) 
and Faddeev's equations can, at least in principle, be solved by 
(numerical) iteration\cite{Merkuriev:1993a}.

The two-body transition matrix is obtained by inverting the Faddeev's 
equation in Eq.~(\ref{fadeqn}) to yield
\begin{equation}
\tilde{\bf T}_{12} 
= \left[ \begin{array}{cc} X_{12} & 0 \\ 0 & X_{21} \end{array} \right]
\left[ \begin{array}{cc} \tilde T_1 & -\tilde T_1\tilde T_2 \\
-\tilde T_2\tilde T_1 & \tilde T_2 \end{array}\right],
\label{12-T12}
\end{equation}
where the $X_{ij}$'s are solutions to the integral equations,
\begin{equation}
\big[ 1-\tilde T_i\tilde T_j\big]X_{ij}=1.
\label{def-Xij}
\end{equation}
The corresponding three-body transition matrix is
\begin{equation}
\tilde{\bf T}_{123} =
\left[ \begin{array}{ccc}
X_{1[23]} & 0 & 0 \\[1mm] 0 & X_{2[31]} & 0 \\[1mm] 0 & 0 & X_{3[12]} 
\end{array}\right]
\left[ \begin{array}{ccc}
\tilde T_1 & -\tilde T_1\tilde G_3\tilde T_2X_{32} 
& -\tilde T_1\tilde G_2\tilde T_3X_{23} \\[1mm]
-\tilde T_2\tilde G_3\tilde T_1X_{31} & \tilde T_2 
& -\tilde T_2\tilde G_1\tilde T_3X_{13} \\[1mm]
-\tilde T_3\tilde G_2\tilde T_1X_{21} 
& -\tilde T_3\tilde G_1\tilde T_2X_{12} & \tilde T_3
\end{array}\right],
\label{123-rtr}
\end{equation}
where the three-body effective Green's functions, $X_{i[jk]}$, 
($i\neq j\neq k$), satisfy the equation
\begin{equation}
X_{i[jk]} \big[ 1 -\tilde T_i\tilde G_j\tilde T_k X_{jk}
-\tilde T_i\tilde G_k\tilde T_j X_{kj} \big] = 1.
\label{Xijk-def}
\end{equation}
We refer to Ref.~\refcite{Shajesh:2011ec} for further details and
expressions for irreducible transition matrices.

%--------------------------------------------

\section{Casimir energies for parallel semitransparent $\delta$-plates}

For scalar fields semitransparent plates are
described by $\delta$-function potentials
\begin{equation}
V_i({\bf x}) = \lambda_i \delta(z-a_i),
\label{Vi-def}
\end{equation}
where $a_i$ specifies the position of the $i$-th plate on the $z$-axis, and
$\lambda_i>0$ is the coupling parameter. In the strong coupling limit, 
$\lambda_i\to\infty$, the potential of Eq.~(\ref{Vi-def})
simulates a plate with Dirichlet boundary conditions.
The total energy $E_{i}$ for a single semitransparent plate is
\begin{equation}
E_{i}(\lambda_i) 
= E_0 + \Delta E_i(\lambda_i)
\label{Ei}
\end{equation}
and the total energy $E_{12}$ of two parallel semitransparent plates 
may be decomposed as,
\begin{equation}
E_{12}(\lambda_1,\lambda_2,{a_{12}}) 
= E_0 + \Delta E_1(\lambda_1) + \Delta E_2(\lambda_2) 
+ \Delta E_{12}(\lambda_1,\lambda_2,{a_{12}}),
\label{E12-2}
\end{equation}
where $a_{12}$ is the distance between the two plates and
\begin{eqnarray}
\frac{E_0}{(\text{Vol})} &=& - \frac{1}{12\pi^2}\int_0^\infty\kappa^3 d\kappa,
\\ 
\frac{\Delta E_i}{L_xL_y}
&=&+\frac{1}{12\pi^2}\int_0^\infty\kappa^2 d\kappa\,\bar{t}_i
{\xrightarrow{{\scriptstyle\lambda_i\to\infty}} }
+\frac{1}{12\pi^2}\int_0^\infty\kappa^2 d\kappa,
\\ 
\frac{\Delta E_{ij}}{L_xL_y} 
&=& - \frac{1}{12\pi^2} \int_0^\infty \kappa^2d\kappa
\frac{ \big[ 2\kappa {a_{ij}} + (1-\bar{t}_i) + (1-\bar{t}_j) \big] }
{ \left[ \frac{1}{\Delta_{ij}} -1 \right]^{-1} }
\xrightarrow{{\scriptstyle\lambda_i\to\infty}}
-\frac{\pi^2}{1440}\frac{1}{a_{ij}^3}.
\label{2-semi} 
\end{eqnarray}
The single-body dimensionally reduced transition matrix and the 
two-body determinants are
\begin{equation}
\bar{t}_i = \frac{\lambda_i}{\lambda_i + 2\kappa}
\xrightarrow{{\scriptstyle\lambda_i\to\infty}}1,
\qquad \Delta_{ij} = 1 - \bar{t}_i \bar{t}_j e^{-2\kappa {a_{ij}}}
\xrightarrow{{\scriptstyle\lambda_i\to\infty}}
(1-e^{-2\kappa {a_{ij}}}).~~ 
\end{equation}
The Casimir energy for free space $E_0$ is divergent and not well defined,
but the trace-log formula formally gives a negative expression.
For a single plate the irreducible single-body Casimir energy also is
divergent and the corresponding expression is positive.
The irreducible two-body Casimir energy is unambiguously finite and negative.
The above expressions also gives the same behavior in the Dirichlet limit.

It is instructive to analyze the two-body contribution to the the 
energy in the limit $a_{12}\to 0$. In this limit the two plates
overlap and we are interested in the distinction between 
a single plate versus two overlapping plates.
Two overlapping plates treated as a single body have the vacuum energy
\begin{equation}
E_{(1+2)}(\lambda_1+\lambda_2) = E_0 + \Delta E_{(1+2)}(\lambda_1+\lambda_2).
\end{equation}
Comparing with the same vacuum energy,
given by Eq.~(\ref{E12-2}) in this limit, expressed in terms of 
irreducible one--and two-body contributions, we identify
\begin{equation}
\Delta E_{12}(\lambda_1,\lambda_2;a_{12}\to 0)
= \Delta E_{(1+2)}(\lambda_1+\lambda_2) 
- \Delta E_1(\lambda_1) - \Delta E_2(\lambda_2),
\label{E12=difsum}
\end{equation}
where the $(1+2)$ in the subscript denotes the two overlapping plates. 
In the limit $a_{12}\to 0$ the irreducible two-body Casimir energy thus
is formally a (divergent) difference of single-body Casimir energies.
The above analysis and conclusions survive in the Dirichlet limit on 
either plate or on both plates, if
the limit $a_{12} \to 0$ is taken before the strong coupling limit.
Although Eq.~(\ref{E12=difsum}) compares divergent expressions
we will see in the following that the limit of two coinciding plates
is well defined for the irreducible three-body Casimir energy.

%----------
\begin{center}
\begin{figure}
\includegraphics[scale=1.0]{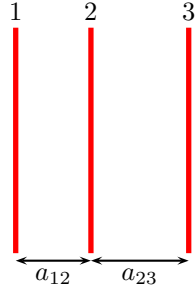}
\caption{Three parallel plates. Plates $1$ and $3$ are separated by
distances $a_{12}$ and $a_{23}$ from the center plate $2$.}
\label{3paraPl}
\end{figure}
\end{center}
%----------

The total Casimir energy for three parallel semitransparent plates is 
decomposed in terms of the irreducible many-body Casimir energies as
\begin{equation}
E_{123} = E_0 + \Delta E_1 + \Delta E_2 + \Delta E_3
+ \Delta E_{12} + \Delta E_{23} + \Delta E_{31} + \Delta E_{123},
\label{E-123-ser}
\end{equation}
where the parameter dependences have been suppressed. (See Fig.~\ref{3paraPl}.)
$\Delta E_{123}$ of three parallel semitransparent plates was
obtained in Ref. \refcite{Shajesh:2011ec} which is finite and positive.
In the Dirichlet limit for all three plates the irreducible three-body
Casimir energy cancels the well-known two-body interaction between the
outer Dirichlet plates,
\begin{equation}
\frac{\Delta E_{123}^D}{L_xL_y} =\frac{\pi^2}{1440}\frac{1}{a_{13}^3}
= -\frac{\Delta E_{13}^D}{L_xL_y},
\end{equation}
where $a_{13}$ is the distance between the outer two plates.
The previous analysis of overlapping plates can be extended to three plates 
when two of the plates overlap. In this case we find
\begin{eqnarray}
\Delta E_{123}(\lambda_1,\lambda_2,\lambda_3;a_{12}\to 0,a_{13})
&=& \Delta E_{(1+2)3}(\lambda_1+\lambda_2,\lambda_3;a_{13})
\nonumber \\ &&
- \Delta E_{13}(\lambda_1,\lambda_3;a_{13})
- \Delta E_{23}(\lambda_2,\lambda_3;a_{23}).  \hspace{5mm}
\end{eqnarray}
Thus, remarkably, in the limit when two plates overlap,
the irreducible three-body contribution can be written  as a difference
of finite irreducible two-body contributions.
The above conclusion survives the strong coupling limit if 
the limit of overlapping plates is taken before the Dirichlet limit. 

%-----------------------------------------------------
\section{(Scalar) atom-like localized potentials above a Dirichlet plate}

In Refs.~\refcite{Rodriguez:2009fn} and \refcite{Varela:2011fe}
the configuration of two spheres above a surface was considered.
We here investigate the scalar analog which is further simplified 
by considering atom-like potentials. The analogous interaction of
atoms on a dielectric plate was explored in Ref.~\refcite{Shajesh:2011md}.
The following scalar analysis might explain why
certain bonds between molecules are weakened in the presence of a metal sheet.

\begin{center}
\begin{figure}
\includegraphics[width=60mm]
{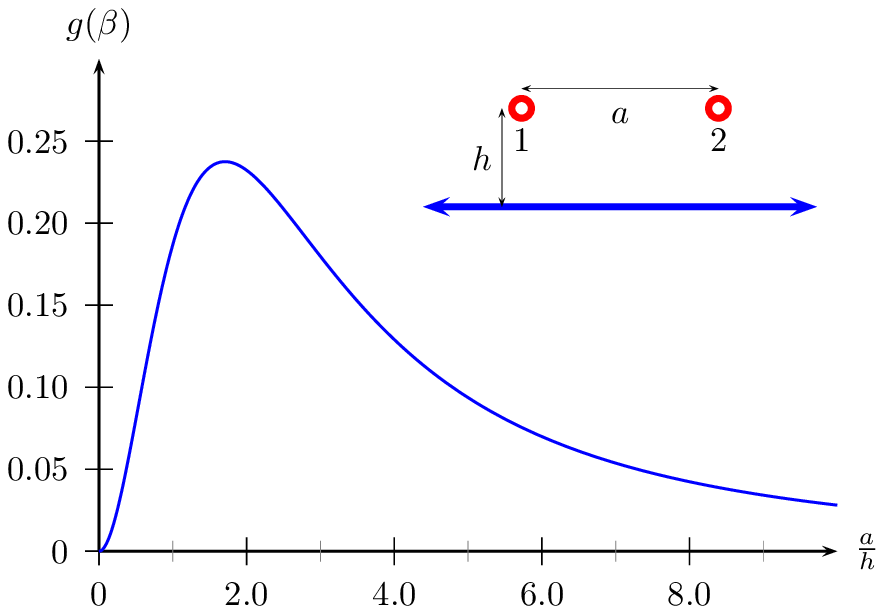}
\hspace{5mm}
\includegraphics[width=60mm]
{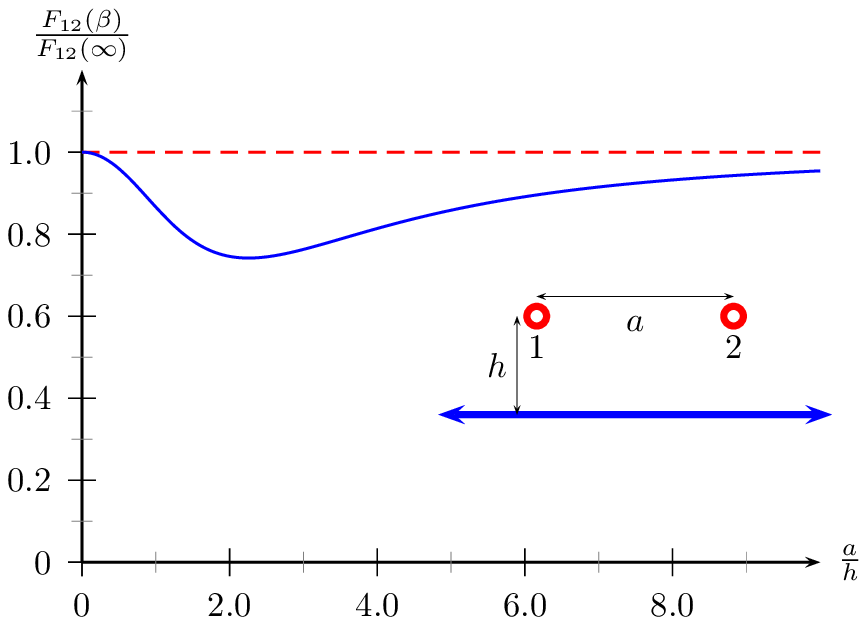}
\caption{Two atom-like potentials above a Dirichlet plate.}
\label{atom-cat-dis-fig}
\end{figure}
\end{center}

Scalar atom-like potentials will be described by
\begin{equation}
V_i({\bf x})=\lambda_i\,\delta^{(3)}({\bf x}-{\bf x}_i),
\qquad i=1,2,
\end{equation}
where $\lambda_i$ now has dimensions of length and ${\bf x}_i$ gives 
the position of the individual atom. 
We chose the two atoms to be at the same height $h$ above the Dirichlet plate
and separated by distance $a$. (See Fig.~\ref{atom-cat-dis-fig}.)
The expressions for two-body and three-body Casimir energies in 
Ref.~\refcite{Shajesh:2011ec} in this case lead to
\begin{equation}
\Delta E_{12} = - \frac{\lambda_1\lambda_2}{64\pi^3a^3},
\qquad \Delta E_{i3} = - \frac{\lambda_i}{32\pi^2h^2},
\qquad \Delta E_{123} = +\frac{\lambda_1\lambda_2}{64\pi^3a^3}\; g(\beta),
\end{equation}
where
\begin{equation}
g(\beta) = \frac{2}{\beta(1+\beta)} -\frac{1}{\beta^3},
\quad \text{with} \quad \beta=\sqrt{1+\left(\frac{2h}{a} \right)^2}.
\end{equation}
The correction factor to the interaction energy between the two
atom-like potentials is given by $g(\beta)$ and is
plotted with respect to the ratio $a/h$ in Fig.~\ref{atom-cat-dis-fig}.
The corresponding force between our scalar atoms is
\begin{equation}
{\bf F}_{12} =
- \frac{3\lambda_1\lambda_2}{64\pi^3 a^4} 
\left[ 1-\Delta_a \right] \hat{\bf a}
- \frac{(\lambda_1+\lambda_2)}{32\pi^2 h^3} 
\left[ 1-\Delta_h \right] \hat{\bf h},
\end{equation} 
where
\begin{eqnarray}
\Delta_a &=& \frac{2}{3} \frac{1}{\beta(1+\beta)}
\left[ 1 + \frac{1}{\beta^2} + \frac{1}{(1+\beta)} + \frac{1}{\beta(1+\beta)}
\right] - \frac{1}{\beta^5}, \\
\Delta_h &=& \frac{1}{\pi}
\frac{1}{ \left( \frac{h}{\lambda_1} + \frac{h}{\lambda_1} \right)}
\frac{h^5}{a^5} \frac{1}{\beta^2}
\left[ \frac{2}{(1+\beta)} \left( \frac{1}{\beta} + \frac{1}{(1+\beta)} \right)
 - \frac{3}{\beta^3} \right].
\end{eqnarray}
The correction to the force-component in the 
direction of $\hat{\bf a}$ is plotted in Fig.~\ref{atom-cat-dis-fig}
and around $a\sim 2h$ reduces the attraction between the scalar atoms
by more than 20\%.

%--------------------------------------------

\section{Weak potentials interacting with a Dirichlet plate}

Let us next consider the more general case of
two independent potentials, $V_i({\bf x})$, $i=1,2$,
describing two objects interacting weakly with each other and 
with a Dirichlet plate placed at $z=0$, 
$V_3=\lambda_3 \delta (z-a_3)$, with $\lambda_3 \to \infty$.
We shall consider only two-dimensional problems and
exclusively deal with potentials that are translationally
symmetric in the $x$-direction. It is convenient to define
Casimir energy per unit length (in $x$-direction),
${\cal E}$, and subtract the single body 
energies and $E_0$ ab initio to write 
\begin{equation}
{\cal E} = \Delta {\cal E}_{12} + \Delta {\cal E}_{23} + \Delta {\cal E}_{31} 
+ \Delta {\cal E}_{123},
\label{E-Eis-E0}
\end{equation}
where ${\cal E}$ on the left hand side is the
total interaction energy per unit length of the three objects.
The two-body Casimir interactions with the Dirichlet plate are given by
the extremely simple expressions\cite{Shajesh:2011ec}
\begin{equation}
\Delta {\cal E}_{i3}^W
= -\frac{1}{32\pi^2} \int_{-\infty}^\infty dy \int_{-\infty}^\infty dz
\frac{V_i(y,z)}{|z|^2},
\qquad (i=1,2),
\label{Ei3DP}
\end{equation}
and the two-body interaction between the two objects
is given by\cite{Milton:2008lw}
\begin{equation}
\Delta {\cal E}_{12}^W
= -\frac{1}{32\pi^3} \int d^2r_1 \int d^2r_2 
\frac{V_1({\bf r}_1) V_2({\bf r}_2)}{r_{12}^2}.
\label{E12v}
\end{equation}
The corresponding three-body contribution to the Casimir energy 
is\cite{Shajesh:2011ec}
\begin{equation}
\Delta {\cal E}_{123}^W
= \frac{1}{32\pi^3} \int d^2r_1 \int d^2r_2 
\frac{V_1({\bf r}_1) V_2({\bf r}_2)}{\bar{r}_{12}^2}
\;Q_3\left(\frac{r_{12}^2}{\bar{r}_{12}^2} \right),
\label{DE123w}
\end{equation}
where the distances are defined as
$r_{12}^2=(y_1-y_2)^2+(z_1-z_2)^2$ and
$(\bar{r}_{12})^2=(y_1-y_2)^2+(|z_1|+|z_2|)^2$,
and the kernel $Q_3$ is
\begin{equation}
Q_3(x) = -\frac{2\ln x}{(1-x)}-1.
\label{Q3k}
\end{equation}

\begin{center}
\begin{figure}
\includegraphics[width=6cm]{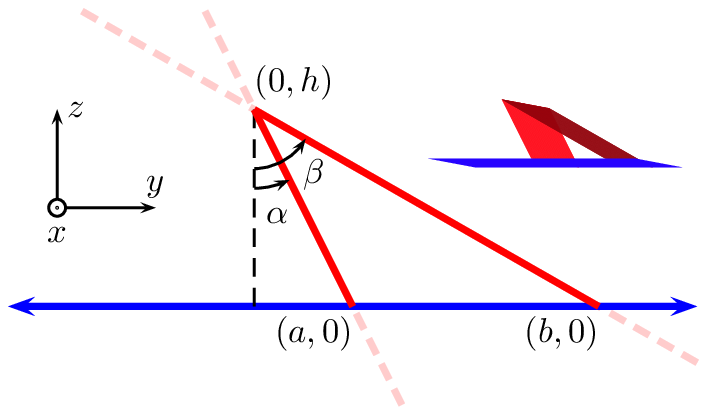}
\hspace{5mm}
\includegraphics[width=6cm]{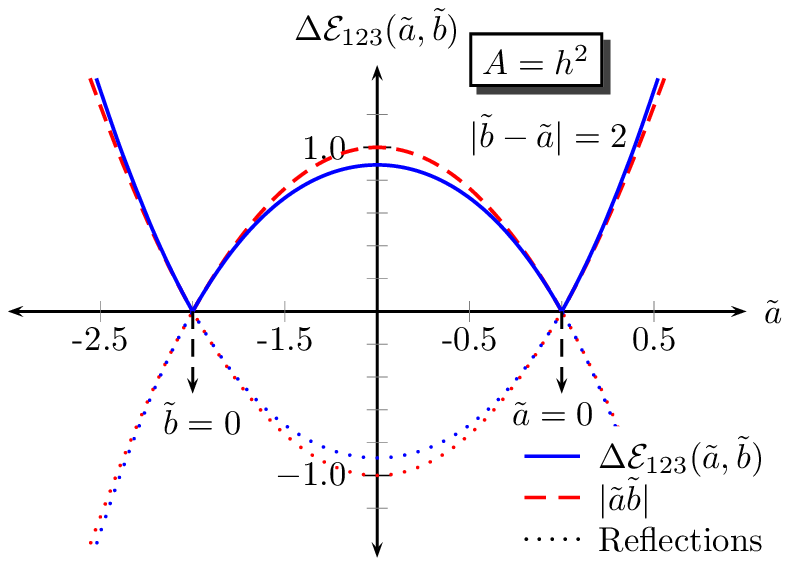}
\caption{Left: Weakly interacting triangular-wedge on a Dirichlet plate.
The objects are of infinite extent in the $x$-direction.
The weakly interacting sides of the wedge (in red) have finite length.
Right: 
$\Delta{\cal E}_{123}(\tilde a,\tilde b)$
as a function of  $\tilde a$ for fixed area,
$A=h^2$. The irreducible three-body Casimir energy is minimal when
the shorter side of the wedge is perpendicular to the Dirichlet plate
($\tilde a=0$ or $\tilde b=0$).
The maximum in the intermediate region corresponds to the unstable
equilibrium of an isosceles triangle. The dashed curves are the 
approximation $\Delta{\cal E}_{123}(\tilde a,\tilde b)\sim |\tilde a\tilde b|$.
The dotted curves are reflections about the $\Delta{\cal E}_{123}=0$ line.
}
\label{wedoTri}
\end{figure}
\end{center}

In Ref.~\refcite{Shajesh:2011ec}
we considered a triangular-wedge with two sides 
described by weak potentials atop a Dirichlet plate at $z=0$, 
forming a waveguide of triangular cross-section:
\begin{subequations}
\begin{eqnarray}
V_1(y,z) &=& \lambda_1 \delta (-z+m_\alpha(y-a)) \,\theta_1, \\
V_2(y,z) &=& \lambda_2 \delta (-z+m_\beta(y-b)) \,\theta_2, \\
V_3(z) &=& \lambda_3 \delta (z), 
\quad \text{with} \quad \lambda_3\to\infty,
\end{eqnarray}
\label{wed-pot-w}%
\end{subequations}
where $\theta_1\equiv\theta(y-\text{min}[0,a])\,\theta(\text{max}[0,a]-y)$
and $\theta_2\equiv\theta(y-\text{min}[0,b])\,\theta(\text{max}[0,b]-y)$.
The sides of the wedge have
lengths $\sqrt{h^2+a^2}$ and $\sqrt{h^2+b^2}$.
The constraint $m_\alpha a=m_\beta b=-h$ forces the sides to intersect 
at $(y=0,z=h)$, where $h$ is the height of the triangle. 
The base of the triangle formed then measures $|b-a|$.
Note that the Dirichlet plate at $z=0$ is of infinite extent.
This triangular-wedge on a Dirichlet plate is depicted in FIG.~\ref{wedoTri}.
Suitable parameters for describing the triangular waveguide are 
$(h,\tilde{a}=a/h,\tilde{b}=b/h)$.
Without loss of generality we measure all lengths in multiples of
the height $h$. The triangle then has height $h=1$ and
the parameter space for the triangle is $-\infty<a,b<\infty$.

%---------
%---------

In FIG.~\ref{wedoTri} we plot the three-body energy for the above configuration
as a function of $\tilde a$ for fixed
area: $A=h^2$, or $|\tilde b - \tilde a|=2$.
The three-body Casimir energy for a waveguide of given cross-sectional
area is minimal
when the shorter side of the wedge is perpendicular to the Dirichlet plate
($\tilde a=0$ or $\tilde b=0$). 
In the intermediate region the energy is extremal for an isosceles
triangle $(-\tilde a =\tilde b=1$) with 
$\Delta{\cal E}_{123}(-1,1)=0.893112\ldots$.
The dashed curve in FIG.~\ref{wedoTri} represents the approximation 
$\Delta{\cal E}_{123}(\tilde a, \tilde b)\sim |\tilde a\tilde b|$.
Remarkably, this extremely simple expression for
the irreducible three-body energy is accurate to better than ten percent
everywhere.

A similar configuration involving parabolic surfaces was also 
analyzed in Ref.~\refcite{Shajesh:2011ec}. No qualitative change in the
three-body Casimir energy was observed for parabolic surfaces.

%-----------------------------------------------------
\section{Future extensions}

For scalar case
we have shown that irreducible three-body parts of Casimir energies can 
contribute significantly. The formalism and 
expressions for the many-body Green's functions and Casimir energies
readily generalize to the electromagnetic case and
we intend to study two real atoms above a metal surface
using the methods developed in Refs.~\refcite{Shajesh:2011ec} and
\refcite{Shajesh:2011md}.
We intend to check if three-body effects contribute significantly 
in the electromagnetic case.

%-----------------------------------------------------

\section*{Acknowledgments}

We would like to thank the organizers of QFEXT11 for a very 
productive workshop.
This work was supported by the National Science Foundation with 
Grant no.~PHY0902054.

%-----------------------------------------------------
\bibliographystyle{unsrt}
\bibliography{biblio/b20110120-qfext11}
%\nocite{*} %%% Will print the complete bib data.
%-----------------------------------------------------

\end{document}